\documentclass[sigconf]{acmart}

\usepackage{color,soul}
\usepackage{makecell}
\usepackage{enumitem}

\usepackage[compact]{titlesec}

\titlespacing{\section}{0pt}{2ex}{1ex}
\titlespacing{\subsection}{0pt}{1ex}{0ex}
\titlespacing{\subsubsection}{0pt}{0.5ex}{0ex}

\setlist{noitemsep, topsep=0pt, parsep=0pt, partopsep=0pt, leftmargin=*}

\renewenvironment{thebibliography}[1]{%
	\begin{oldthebibliography}{#1}%
		\setlength{\parskip}{0ex}%
		\setlength{\itemsep}{0ex}%
	}%
	{%
	\end{oldthebibliography}%
}

\AtBeginDocument{%
	\providecommand\BibTeX{{%
			\normalfont B\kern-0.5em{\scshape i\kern-0.25em b}\kern-0.8em\TeX}}}

 \renewcommand\footnotetextcopyrightpermission[1]{} 

\begin{document}
	

	\title{The Final Frontier: Deep Learning in Space
	}

	\author{%
		Vivek Kothari$^{\dagger \ast}$, Edgar Liberis$^{\dagger \ast}$, Nicholas D.~Lane$^{\dagger \diamond}$  
	}
	
	
	\affiliation{\institution{$^\dagger$University of Oxford\hspace{+0.75cm}$^\diamond$Samsung AI}}

	
	
	
	
	\begin{abstract}
		Machine learning, particularly deep learning, is being increasing utilised in space applications, mirroring the groundbreaking success in many earthbound problems. Deploying a space device, \emph{e.g.} a satellite, is becoming more accessible to small actors due to the development of modular satellites and commercial space launches, which fuels further growth of this area. Deep learning's ability to deliver sophisticated computational intelligence makes it an attractive option to facilitate various tasks on space devices and reduce operational costs. In this work, we identify deep learning in space as one of development directions for mobile and embedded machine learning. We collate various applications of machine learning to space data, such as satellite imaging, and describe how on-device deep learning can meaningfully improve the operation of a spacecraft, such as by reducing communication costs or facilitating navigation. We detail and contextualise compute platform of satellites and draw parallels with embedded systems and current research in deep learning for resource-constrained environments. 
	\end{abstract}

	\pagestyle{empty}
	
	\maketitle
	
	\section{Introduction}
	
	Machine learning scales to and thrives in data-abundant environments making it well suited to applications in space. Satellite imagery is ubiquitous in space. Produced by both imaging the Earth and space (telescope satellites), it allows learned models to power a range of monitoring tasks. Machine learning can also play an active role in the operation of a spacecraft, allowing for precise automated control and facilitating on-board tasks, such as docking or navigation.
	
	
	Machine learning's impact in space applications will continue to grow as cheaper satellite platforms mature and become more accessible to small actors, thus widening the range of possible activities in space. Similarly, hardware innovations from terrestrial systems, such as the multi-core design and specially-designed accelerators, are increasing the compute power available to spacecraft.
	
	
	
	This increased accessibility of space platforms and space data is fuelling excitement in ``space activity'' as a new direction for applied machine learning researchers with potentially transformative results for future applications and satellite hardware. In particular, recent advances in machine learning (ML) and deep learning (DL) in constrained environments~\cite{liberis2019neural, chowdhery2019visual, wang2018towards, sze2017efficient} would enable running neural networks on the spacecraft itself. Doing so will enable many ``smart'' applications to be run in space autonomously (as the communication channel between the spacecraft and ground station is often limited), which is likely to have a similar impact and development trajectory to that of terrestrial smart devices, homes, embedded systems and mobile phones~\cite{lane2017squeezing}.
	
	In addition to challenges faced by DL for embedded systems, space imposes extra requirements: the need for radiation-hardened hardware, robustness, and extensive verification. We discuss those challenges, demonstrating how space is a similar and yet unique environment for DL applications, which makes applying developments from embedded systems non-trivial.
	
	In this work, we argue that ML/DL in space is an important direction for mobile and embedded computing (MEC) moving forward. We summarise prominent applications of ML in space, give an indication of the range of compute and sensor capabilities of space hardware, and present pointers for future work, thus providing the reader with the necessary information to start exploring this area. We quantitatively illustrate how DL based solutions can offer double the power efficiency compared to the current state of the art.
	\footnote{A version of this paper was accepted at HotMobile 2020: The 21st International Workshop on Mobile Computing Systems and Applications.}

	\section{Deep Learning Meets Space}
	\label{sec:applications}
	
	
	Spacecraft (vehicles designed for operation outside the earth's atmosphere) and satellites (objects that orbit a natural body) have two types of systems: payload, which comprises instruments that faciliate the primary purpose of the spacecrafts; and operations systems, which support the payload and allow it to reach, stay, and work in space. This section will analyse current applications of ML pertaining to both systems.  
	
	\subsection{Analysis of Payload Data}
	Earth observation satellites in either geostationary (GEO) or Low Earth (LEO) orbits carry sensors ranging from RGB imagers for cloud cover detection to more
	specialised sensors for other atmospheric properties: temperature, humidity, wind vectors, and gaseous composition. Data, often from radiometric or spectral sensors, has traditionally been processed at ground station primarily using classical ML and hand-crafted specialised algorithms. 
	
	The similarity of terrestrial and space sensor modalities makes ML/DL well-suited for payload data. The following areas are only a few that have seen remarkable success from such methods. We will later (Sec.~\ref{sec:caseImaging}) quantitatively demonstrate how DL methods can help save power on satellites.
	
	
	\textbf{Weather \& atmospheric monitoring.} Cloud detection \cite{lewis1997determination} and estimating precipitation~\cite{ba2001goes}, and green-house gas concentrations use classical ML.\cite{wimmers2019using} use Faster-RCNNs to achieve remarkable results in estimating tropical storm intensity, but the required data is limited and has to be aligned to microwave wavelengths.
	
	\textbf{Vegetation and ground cover classification.} Hyperspectral data (HSD) is used to identify land cover ~\cite{baker2007change}, with data from MODIS and LANDSAT satellites successfully used to show the diminishing wetlands. 
	
	DL models naturally excel at challenges presented by the HSD. In a hyperspectral image, a pixel represents a large spatial area which may have several types of vegetation. This leads to the entanglement of spectral signatures and, when compounded with the high-dimensionality and a large intra-class variability of HSD, makes ground cover classification a formidable problem. 3D-CNNs and ResNets ~\cite{manning2018machine} successfully address these challenges on ISS data to achieve 96.4\% classification accuracy. \cite{paoletti2019deep} review a number of DL architectures and demonstrate their effectiveness in situations with semi-supervised or sparse data.
	
	
	
	\textbf{Object detection and tracking.} When pointed towards earth, HSD has been used to detect humans during natural disasters, track endangered animals~\cite{guirado2019whale}, military troops and ships and monitor oil spills~\cite{salem2001hyperspectral}. When pointed at the sky, they leverage the lack of atmospheric interference to detect galactic phenomena. The James Webb Space Telescope project is one of the first to use DL in data post-processing to detect galaxy clusters~\cite{chan2019deep}.

	\subsection{General Spacecraft Operation}
	Operation systems include the Guidance, Navigation and Control (GNC), communication, power, and propulsion systems. NASA has defined four levels of spacecraft autonomy~\cite{drabbe_drabbe_2008}. The lowest level corresponds to a primarily ground controlled mission, whereas the highest expects an ability to independently re-evaluate goals. Currently automation is provided by pervasive \textit{on-board control procedures (OBCPs)}. Used in satellites such as the Rosetta, Venus, Herschel \& Planck~\cite{ferraguto2008board}, OBCPs initiate a predefined series of actions when an event is detected. DL can improve not only OBCPs, through better event detection and subsequent planning, but also systems which follow.
	
	\textbf{Communication.} Software-defined radio (SDR) is replacing multiple antennae designs. Its communication protocols can depend on many parameters, such as packet re-transmission rate and band. The use of DL, such as reinforcement learning (RL) to dynamically optimise parameters has been proposed for space applications but not yet adopted~\cite{ferreira2019reinforcement, ortizuse}.
	
	\textbf{Automated control and navigation.} The GNC and propulsion systems control the movement of a spacecraft. Historically the majority of these manoeuvring operations were directed by a human, which is cumbersome and only feasible for near Earth missions. Onboard DL systems would bring a much needed degree of autonomy and robustness to GNCs, which is particularly important for deep space missions that suffer from lag and gaps in communication. 
	
	
	Positioning becomes important during docking or landing. Typically, such systems almost exclusively rely on LIDAR, but newer techniques, such as the natural feature tracking (NFT), use optical systems. \citet{evers2019deep} uses the \textit{YOLO} vision model to estimate pose and relative distance, which achieves a 98\% accuracy on the author's dataset and lays the groundwork for real-time models. Despite impressive accuracy, DL systems would require large banks of images to train on~\cite{lorenz2017lessons}.
	
	Spacecraft are particularly sensitive during docking and landing: several tonnes need to be moved with \textit{centimetre-level} precision. While traditionally the domain of control systems, preliminary work applies reinforcement learning methods to 6-DOF cold gas thruster systems~\cite{nanjangud2018robotics}.  
	
	While spacecraft travel space, Landers and rovers must traverse non-Earth surfaces. While they do not use deep models, conventional ML/control systems like AEGIS have been used on the MER project~\cite{estlin2012aegis}. Projects like the Surrey Rover Autonomy Software \& Hardware Testbed (SMART)~\cite{gao2012modular} provide terrestrial simulation facilities and are looking at modular (deep) systems. \citet{blacker2019rapid} use a yet-to-be-deployed CNN based system which judges the navigability of each part of the terrain then plans a safe path based on the results. The system can be tuned to run at different latency and memory capacities. GNC's similarity to terrestrial problems of vision and autonomous driving make it a particularly attractive area of development.
	
	\section{Space Hardware and Software}
	\label{sec:hardNSoftware}
	
	Computational resources in space have traditionally been highly specialised, tightly-integrated monoliths. In contrast with terrestrial hardware systems, the harsh and remote environment of space requires compute systems (incl. the processor and memory chips) to be simultaneously efficient, radiation-resistant, and fault-tolerant. In addition systems sent into space have to be thoroughly verified. As a result, space systems, especially hardware, lag considerably behind modern compute. 
	
	\subsection{System Platforms}
	
	Spacecraft have highly mission-dependent designs (Tab.~\ref{tbl:hardware}), with purpose-designed hardware (ASICs) or high end micro-controllers powering older space missions. However, such systems are high cost, non-resilient, and large. With time, these specialisations became infeasible along several axes (power, cost, weight, volume), thus to reduce the development costs, systems are increasingly being assembled using off-the-shelf components (COTS). This applies to both smaller spacecraft, such as CubeSats, and large multi-million dollar satellites.
	
	
	
	\textbf{System Software and Operating systems.} 
	Due to their specificity, historically, large satellites have had minimal software interfaces: either embedded modules, like vxWorks RTEMS, or purpose-made minimal software with a tightly coupled software/hardware interface~\cite{torelli_2019}. These systems varied dramatically and often did not support floating-point calculations, integer multiplication or division units, or interrupt or dynamic memory allocation. Contemporary embedded systems still in use include the core Flight System (cFS) and core Executive (cFE) from Goddard Space Flight Centre, and COSMOS by Ball Aerospace.
	Projects like, such as SPINAS \cite{notebaert}, are attempting to create a more uniform (equal-capability), and open source, platform for smaller spacecraft with more sophisticated compute. Some newer spacecraft also see Linux-based OSes~\cite{nasasmallsat}.
	
	
	
	\textbf{Memory and Compute Capabilities.} 
	Modern space compute systems are moving towards shared/re-configurable~\cite{george2018onboard}, multi-core systems which would be capable of running DL models. For example, the recent JUICE mission to map Jupiter's moons uses a common digital processing unit (DPU) and software packages, which are shared between 10 of its instruments~\cite{torelli_2019}. Other systems may utilise multicore processors, \emph{e.g.} LEON-GR740 (32 bit, quad-core, rad-hard SOC), with cores preferentially being assigned to specific tasks (such as navigation). Recently, some workloads, such as linear algebra, are being accelerated with field programmable gate arrays (FPGAs) or low-power GPU-like accelerators~\cite{notebaert}. Special low-power accelerators can be incorporated into an FPGA-based on-board computer or as a separate chip, \emph{e.g.} the Movidius compute stick, which was being tested for deployment in space. Tab.~\ref{tbl:hardware} shows some of these processing elements and their specifications.
	
	
	\begin{table*}[]
		\small
		\begin{tabular}{lrrr}
			\hline
			\textbf{Type} & \textbf{Compute platform} & \textbf{Specifications} & \textbf{Power budget} \\ \hline
			microcontroller & TI MSP430F2618 & 12MHz, 8KB SRAM, 116KB FLASH, X-band (8kbps) & 35W \\
			processor SOC & BAE RAD 750 & 200MHz, 2GB flash 256MB DRAM  & 5W \\
			accelerator & Intel Movidius NCS & VPU, 4GB RAM & 1W \\
			microcontroller & VA41630 (Cortex M4) & 100 MHz, 64KB SRAM, 256KB FLASH & -- \\
			FGPA & Xilinx Virtex-5QV & 81920 LUT6, 596 RAMB16, 320 DSPs, 65 nm SRAM & 5–10 W \\ \hline
		\end{tabular}%
		
		\caption{\textmd{Different types of space hardware and their configurations. The 1st entry was used on a Mars Cube One (CubeSat) mission, while 2nd is charted to be used on the Mars 2020 interplanetary rover. The power envelope of these devices is magnitudes lower, compared to the WorldView 3 imaging satellite (3100 W).}}
		\label{tbl:hardware}
		\vspace{-0.5cm}
	\end{table*}

	\textbf{Power budget.}
	The largest limiting factor for on-board compute is power. Power generation, storage, and dissemination is facilitated through the electrical power systems (EPS). The wattage of supplied power is typically adjusted for payload requirements, which ranges 20W to 95W. In small satellites, power is generated through multi-junction solar cells, which have 28-38\% efficiency and so need to be quite large to sustain the required power output. Power is most commonly stored in rechargeable Li-ion or Li-Po batteries ranging from 58-243 Wh/Kg~\cite{nasasmallsat}.

	\subsection{Radiation Hardening}
	In space, devices are no longer protected from Sun's radiation by the Earth's atmosphere, which can cause spurious errors or stuck transistors in the device's circuitry. Radiation damages the hardware either through its cumulative effects (total ionizing dose, TID) or through single event effects (SEE). Recoverable SEEs are called single event upsets (SEU) and can affect the logic state of memory. Radiation hardening (rad-hard) allows a compute component to withstand such errors. Rad-hard components are twice as slow and many times as expensive as their regular counterparts~\cite{gretok2019comparative}. Overheads incurred by the space-grade CPUs are typically much larger than those incurred by the DSP and FPGA because they required more significant decreases in operating frequencies~\cite{gretok2019comparative}.
	
	\textbf{Physical hardening.} This involves using different materials, for example, insulating substrates such as silicon on sapphire~\cite{trivedi2016survey}.Other approaches involve shielding the circuit and alternative doping mechanisms.
	
	\textbf{Circuit based hardening.} This involves adding extra circuitry/logic to correct for the effects of SEEs. These include: watchdog protection, overcurrent circuits, power control and error correcting circuits (\emph{e.g.} CRC and forward error correction in communication boards). Error correction is implemented both at the hardware and software levels, such as in the main memory where hardware level ECC and software EDAC are used in synergy.

	\subsection{Suitability for DL Workloads}
	According to ~\citet{dennehy}, an ML-assisted optical navigation system would have: 2--5 Gbps sensor I/O, 1--10 GOPs CPU, 1GB/s memory bandwidth, 250 Mbps cross link bandwidth to Earth. As the previous sections showed we are almost in a position to put that into a small satellites. 
	
	Space hardware is becoming increasingly closer to the multicore terrestrial mobile edge computing (MEC). While rad-hard components do decrease performance and raise costs, cheaper off-the-shelf components are finding their way to cubesats. Progress in hardware and research in deep learning algorithms for constrained environments have begun to meet at a point which allows deep models to be deployed to space. 
	
	Several DL systems are being initially tested on Earth. \citet{schartel2017increasing} train a SqueezeNet model with the intent of transferring it to a space embedded system; \citet{buonaiuto2017satellite} consider Nvidia-TX1 hardware with the CUDA Deep Neural Network (cuDNN) library and TensorRT software. FPGAs, like the Xilinx Artix-7 and the Xilinx Zynq-7020, have been used for neuromorphic chips and image analysis.
	
	The first space systems are specifically geared towards DL workloads are making their way to production. CloudScout~\cite{esposito_2019}, a cloud identification algorithm, uses a specialised Visual Processing Unit to identify cloud patterns. 
	
	
	\section{Case Study: On-device Satellite Imagery}
	\label{sec:caseImaging}
	
	
	Obtaining satellite imagery is one of the most widespread uses of spaceborne hardware. However sending and receiving large volumes of data is power consuming. In situations, such as with rovers, the high-latency and low bandwidth communication channels make this prohibitive. As seen in section~\ref{sec:hardNSoftware} we are approaching have the hardware, software, and algorithmic capability required to use DL methods on board the highly constrained environment of a spacecraft. In this section, we describe how they can both select relevant imaging data and compress it. We then quantitatively show that DL can save \textit{at least} half the power.
	
	\subsection{Efficient Satellite Imaging}
	
	
	
\begin{figure}
  \centering
  \includegraphics[width=0.22\textwidth]{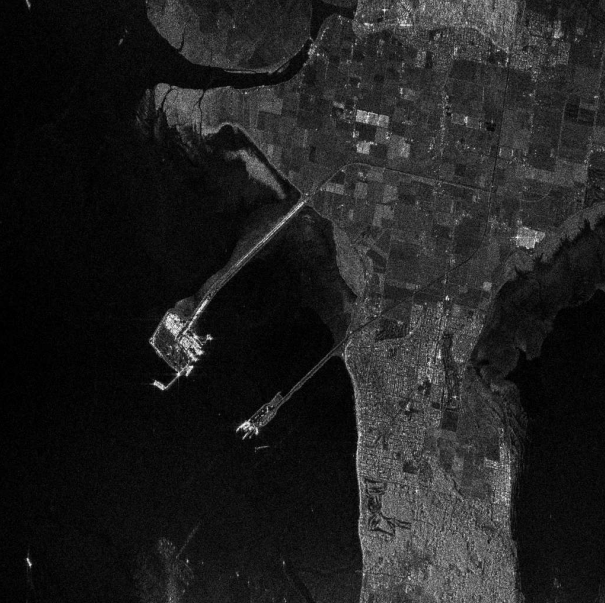}
  \includegraphics[width=0.22\textwidth]{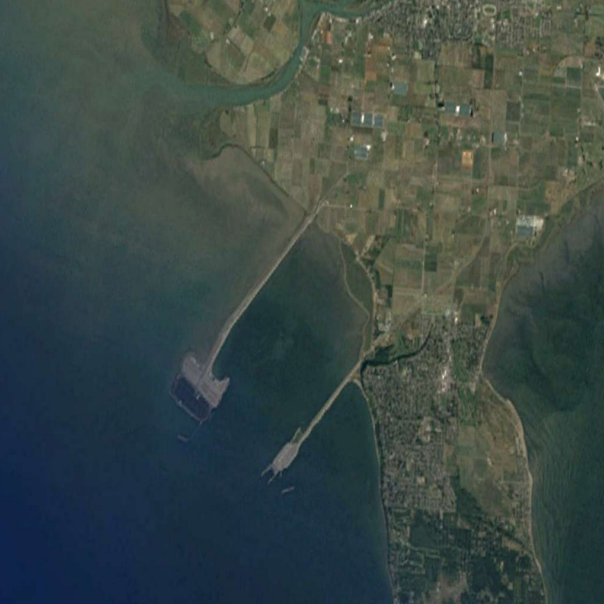}
  \caption{Area imaged using SAR (left) and optical sensors (right). Reproduced from~\citet{wang2018generating}.}
  \label{fig:sar_sat_sample}
\end{figure}
	
	Most imagining sensors capture several bands (commonly between IR to UV) spread across the electromagnetic spectrum. Modern sensors can capture very high resolution (VHR) images at up to 31cm of ground per pixel (in panchromatic mode)~\cite{wv4sat}. Even higher resolution images can be obtained using synthetic-aperture radar (SAR), which uses the motion of radio antenna over the surface to map the surface in three dimensions at a resolution of just a few centimetres per pixel~\cite{sarlecture}. A number of both raw~\cite{usgs} and preprocessed~\cite{38-cloud-1} hyper spectral datasets are readily available.

	
	Captured data needs to be transmitted to the ground station for aggregation and analysis, which can be expensive. A satellite can reduce the amount of data transmitted by employing deep learning: on-board pre-processing can discard parts of the image of no interest, \emph{e.g.} occluded by clouds. Global annual cloud coverage is estimated to be at 66\%, so excluding cloud images would drastically reduce the amount of data transmitted~\cite{jeppesen2019cloud}. For satellites deployed for a particular purpose, such as boat or whale detection, neural networks can also be used to facilitate the primary task of a satellite and only transmit regions of interest. 
	
	
	
	Transmission costs can be further reduced by employing a neural network to compress image data. While the following models offer spectacular gains, training models specifically for satellite data would yield considerably better results. 
	Near-lossless compression~\cite{qian2006near} achieved a 20:1 compression ration(CR) with hyperspectral data. In lossy compression, the Feb 2019 CCSDS standard for on-board lossy compression of hyperspectral images, uses predictive coding on-board and a residual hyperspectral CNN back on Earth to de-quantize the results~\cite{diego2019image} and achieves 0.1 bits per pixel compression ratio, far surpassing classical compression standards, \emph{e.g} JPEG.
	
	\subsection{Potential Efficiency Gains}
	Here we present a typical system and quantify the bare minimum power saving an out-of-the-box DL systems would be able to provide us. Note that this is an approximate calculation: a finely tuned and end-to-end designed system could yield considerably better power gains. 
	
	\textbf{Model.} For example, MobileNet-V2 family of models were shown to run successfully on microcontroller-sized hardware (with 8-bit quantization)~\cite{chowdhery2019visual} and power some image segmentation models (\emph{e.g.} MobileNet-V2-backed DeepLabV3+ model~\cite{chen2018encoder}), which can be used for cloud detection. 
	
	\textbf{Communication.} Assuming the use of an S-band transmitter operating at 13W (for 33dBm output power) with a 4.3Mbps data rate~\cite{sbandantenna}, transferring a 512x512 patch of data at 12-bits per pixel, would take approximately 0.73s and consume around 9.5J of power (ignoring communication protocol overheads).
	
	\textbf{Compute.} If, instead, we perform 3s worth of inference on a LEON3 processor\footnote{Est. under 300MAdds for a MobileNet-V2 on a 512x512 input, running on a 100MHz processor.}~\cite{leon3spec} to achieve at least 20:1 compression ratio (incl. cloud removal), we spend 4.5J on computation and 0.2J on transmission.
	
	\textbf{Power saving.} The above shows a nearly 2x power saving. Considering different model architectures would result in other computation vs transmission power usage trade-offs. 
	
	Performing neural network compression before data transmission can considerably improve transmission latency and power usage, allowing longer mission lengths and the use of less costly transmission hardware. Space devices present an interesting constraint space for ultra-compact computer vision models.

	\section{Challenges and Opportunities Ahead}
	We have barely scratched the surface of the what is possible with deep learning in space. We outline challenges and applications which hold the greatest potential.
	
	
	\subsection{New Applications in Space}
	
	
	Terrestrial DL vision models are built for optical (narrow band) data whereas most image data from space is hyperspectral. Having HSD is especially important because several optical artefacts (\emph{e.g.} metal ground cover, x-ray star bursts) are only often present within a subset of the spectrum and require models capable of searching through deeper data cubes. 
	
	
	
	
	Not only have individual modalities not been completely exploited but due to the lack of compute, multi-modal DL systems have yet to make their way to space. Multimodal approaches would ubiquitously improve spacecraft and payload operation \emph{e.g.} fusing magnetometer, horizon and sun sensor data for GNC operation, and fusing SAR and HSD for terrain characterisation.
	
	There are also numerous applications which are just waiting to see DL methods adapted - Ranging from DL/RL robotic construction in zero gravity environments to DL for crew health monitoring.

	\subsection{Improved Compute Paradigms for Space}
	
	The characterisation of DL models on a spacecraft must encompass more than just accuracy. The ability of the model to perform depends not only its construction but also on an environment constrained in terms of memory, power, compute, and reliability. Thus the definition of efficiency must be expanded to include hardware and context aware characterisation.
	
	The uptake of off-the-shelf components, would allows us to leverage recent developments in compute efficient (quantized)~\cite{sze2017efficient}, memory sensitive (compressed)~\cite{liberis2019neural}, and energy aware ~\cite{wang2018towards} terrestrial DL models. While made significantly easier, the adaption process would still need to accommodate formidable hurdles endemic to space hardware, such as the higher error rates and the increased memory latency with rad-hard components.
	
	Not only must the efficiency of DL models on a compute unit be measured along multiple axes, but it must also be characterised in the context of the overall spacecraft's operation. This becomes increasingly important as compute components in newer spacecraft are shared between various subsystems. Real time systems, \emph{e.g.} navigation, may be sensitive to interrupts and IO bottlenecks. 
	
	As powerful hardware becomes common in space, it may be possible to leverage more than a single satellite for computation. Such networks would offer not only more computational power but also greater fault tolerance.

	

	\subsection{Redefining Robustness and Reliability}
	
	Developing a comprehensive Validation and Verification (V\&V) framework for embedded and robotics systems in critical areas such as healthcare and civil infrastructure is an area of current research. However the remoteness and expense of space make it radically more risk averse and often with higher costs than MEC systems.
	
	
	The current validation standards for space-based systems and software listed in ECSS-E-ST-10-02C Rev.1 and ECSS-E-ST-40C \cite{ECSS_stds} are inadequate for automated DL systems on spacecraft. Core to V\&V is the \textit{qualification} process which has 4 components: analysis, testing, inspection, and demonstration. Each of these components differs significantly when applied to ML systems and especially to DL systems. DL systems are less deterministic, less amenable to isolation and component testing, are data driven, and suffer from a lack of a testing oracle. Adding to the challenge, DL systems in space need to be measured along multiple axes: correctness, robustness, efficiency, interpretability. The rigour and span of V\&V in space set it apart from methods used for MEC systems.
	
	A few preliminary approaches would combine formal verification (analysis) with simulation (testing) and interpretability mechanisms (inspection). Analysis methods utilise sample perturbation or mixed integer linear programming~\cite{duttasherlock} to characterise individual components. \cite{xiang2018verification} surveys the adaptation of more formal verification methods. These methods are only able to characterise a finite set of cases. Inspection of DL using interpretability models such as LIME or contextuality models~\cite{carvalho2019machine} allows for humans-in-the-loop systems both in V\&V and mission control. Finally DL sub-systems are tested through simulations before testing the entire component in the field.
	

	
	
	\section{Conclusion}
	
	As space devices become more affordable to launch and their hardware becomes more powerful to run non-trivial workloads, deep learning in space will continue to grow as a topic within mobile and embedded machine learning. In this work, we presented how machine learning can be applied to space data, drew a parallel between space and terrestrial embedded hardware and showed how on-device deep learning can improve the operation of a space device.
	
	\begin{acks}
		
		This work was supported by the \grantsponsor{EPRSC}{EPRSC}{} through Grants \grantnum{EPRSC}{DTP (EP/R513295/1)} and \grantnum{EPRSC}{MOA (EP/S001530/)}, and \grantsponsor{SAI}{Samsung AI}{}. We would also like to thank Dr. Aakanksha Chowdhery and other anonymous reviewers for their input throughout the submission process for ACM HotMobile 2020.

	\end{acks}
	
	
	\setcitestyle{numbers}
	\small{

\begin{thebibliography}{48}


\ifx \showCODEN    \undefined \def \showCODEN     #1{\unskip}     \fi
\ifx \showDOI      \undefined \def \showDOI       #1{#1}\fi
\ifx \showISBNx    \undefined \def \showISBNx     #1{\unskip}     \fi
\ifx \showISBNxiii \undefined \def \showISBNxiii  #1{\unskip}     \fi
\ifx \showISSN     \undefined \def \showISSN      #1{\unskip}     \fi
\ifx \showLCCN     \undefined \def \showLCCN      #1{\unskip}     \fi
\ifx \shownote     \undefined \def \shownote      #1{#1}          \fi
\ifx \showarticletitle \undefined \def \showarticletitle #1{#1}   \fi
\ifx \showURL      \undefined \def \showURL       {\relax}        \fi
\providecommand\bibfield[2]{#2}
\providecommand\bibinfo[2]{#2}
\providecommand\natexlab[1]{#1}
\providecommand\showeprint[2][]{arXiv:#2}

\bibitem[\protect\citeauthoryear{??}{ECS}{2014}]%
        {ECSS_stds}
 \bibinfo{year}{2014}\natexlab{}.
\newblock \bibinfo{title}{ECSS Active Standards}.
\newblock
\newblock
\urldef\tempurl%
\url{https://ecss.nl/active-standards/}
\showURL{%
\tempurl}


\bibitem[\protect\citeauthoryear{Ba and Gruber}{Ba and Gruber}{2001}]%
        {ba2001goes}
\bibfield{author}{\bibinfo{person}{Mamoudou~B Ba} {and} \bibinfo{person}{Arnold
  Gruber}.} \bibinfo{year}{2001}\natexlab{}.
\newblock \showarticletitle{GOES multispectral rainfall algorithm (GMSRA)}.
\newblock \bibinfo{journal}{\emph{Journal of Applied Meteorology}}
  \bibinfo{volume}{40}, \bibinfo{number}{8} (\bibinfo{year}{2001}),
  \bibinfo{pages}{1500--1514}.
\newblock


\bibitem[\protect\citeauthoryear{Baker, Lawrence, Montagne, and Patten}{Baker
  et~al\mbox{.}}{2007}]%
        {baker2007change}
\bibfield{author}{\bibinfo{person}{Corey Baker}, \bibinfo{person}{Rick~L
  Lawrence}, \bibinfo{person}{Clifford Montagne}, {and} \bibinfo{person}{Duncan
  Patten}.} \bibinfo{year}{2007}\natexlab{}.
\newblock \showarticletitle{Change detection of wetland ecosystems using
  Landsat imagery and change vector analysis}.
\newblock \bibinfo{journal}{\emph{Wetlands}} \bibinfo{volume}{27},
  \bibinfo{number}{3} (\bibinfo{year}{2007}), \bibinfo{pages}{610}.
\newblock


\bibitem[\protect\citeauthoryear{Blacker, Bridges, and Hadfield}{Blacker
  et~al\mbox{.}}{2019}]%
        {blacker2019rapid}
\bibfield{author}{\bibinfo{person}{P Blacker}, \bibinfo{person}{CP Bridges},
  {and} \bibinfo{person}{S Hadfield}.} \bibinfo{year}{2019}\natexlab{}.
\newblock \showarticletitle{Rapid Prototyping of Deep Learning Models on
  Radiation Hardened CPUs}. In \bibinfo{booktitle}{\emph{2019 NASA/ESA
  Conference on Adaptive Hardware and Systems (AHS)}}. IEEE,
  \bibinfo{pages}{25--32}.
\newblock


\bibitem[\protect\citeauthoryear{Buonaiuto, Louie, Aarestad, Mital, Mateik,
  Sivilli, Bhopale, Kief, and Zufelt}{Buonaiuto et~al\mbox{.}}{2017}]%
        {buonaiuto2017satellite}
\bibfield{author}{\bibinfo{person}{Nick Buonaiuto}, \bibinfo{person}{Mark
  Louie}, \bibinfo{person}{Jim Aarestad}, \bibinfo{person}{Rohit Mital},
  \bibinfo{person}{Dennis Mateik}, \bibinfo{person}{Robert Sivilli},
  \bibinfo{person}{Apoorva Bhopale}, \bibinfo{person}{Craig Kief}, {and}
  \bibinfo{person}{Brian Zufelt}.} \bibinfo{year}{2017}\natexlab{}.
\newblock \showarticletitle{Satellite identification imaging for small
  satellites using NVIDIA}.
\newblock  (\bibinfo{year}{2017}).
\newblock


\bibitem[\protect\citeauthoryear{Carvalho, Pereira, and Cardoso}{Carvalho
  et~al\mbox{.}}{2019}]%
        {carvalho2019machine}
\bibfield{author}{\bibinfo{person}{Diogo~V Carvalho},
  \bibinfo{person}{Eduardo~M Pereira}, {and} \bibinfo{person}{Jaime~S
  Cardoso}.} \bibinfo{year}{2019}\natexlab{}.
\newblock \showarticletitle{Machine Learning Interpretability: A Survey on
  Methods and Metrics}.
\newblock \bibinfo{journal}{\emph{Electronics}} \bibinfo{volume}{8},
  \bibinfo{number}{8} (\bibinfo{year}{2019}), \bibinfo{pages}{832}.
\newblock


\bibitem[\protect\citeauthoryear{Chan and Stott}{Chan and Stott}{2019}]%
        {chan2019deep}
\bibfield{author}{\bibinfo{person}{Matthew~C Chan} {and}
  \bibinfo{person}{John~P Stott}.} \bibinfo{year}{2019}\natexlab{}.
\newblock \showarticletitle{Deep-CEE I: Fishing for Galaxy Clusters with Deep
  Neural Nets}.
\newblock \bibinfo{journal}{\emph{arXiv preprint arXiv:1906.08784}}
  (\bibinfo{year}{2019}).
\newblock


\bibitem[\protect\citeauthoryear{Chen, Zhu, Papandreou, Schroff, and Adam}{Chen
  et~al\mbox{.}}{2018}]%
        {chen2018encoder}
\bibfield{author}{\bibinfo{person}{Liang-Chieh Chen}, \bibinfo{person}{Yukun
  Zhu}, \bibinfo{person}{George Papandreou}, \bibinfo{person}{Florian Schroff},
  {and} \bibinfo{person}{Hartwig Adam}.} \bibinfo{year}{2018}\natexlab{}.
\newblock \showarticletitle{Encoder-decoder with atrous separable convolution
  for semantic image segmentation}. In \bibinfo{booktitle}{\emph{Proceedings of
  the European conference on computer vision (ECCV)}}.
  \bibinfo{pages}{801--818}.
\newblock


\bibitem[\protect\citeauthoryear{Chowdhery, Warden, Shlens, Howard, and
  Rhodes}{Chowdhery et~al\mbox{.}}{2019}]%
        {chowdhery2019visual}
\bibfield{author}{\bibinfo{person}{Aakanksha Chowdhery}, \bibinfo{person}{Pete
  Warden}, \bibinfo{person}{Jonathon Shlens}, \bibinfo{person}{Andrew Howard},
  {and} \bibinfo{person}{Rocky Rhodes}.} \bibinfo{year}{2019}\natexlab{}.
\newblock \showarticletitle{{Visual Wake Words Dataset}}.
\newblock  (\bibinfo{year}{2019}).
\newblock
\showeprint[arxiv]{1906.05721}
\urldef\tempurl%
\url{http://arxiv.org/abs/1906.05721}
\showURL{%
\tempurl}


\bibitem[\protect\citeauthoryear{{COBHAM}}{{COBHAM}}{2018}]%
        {leon3spec}
\bibfield{author}{\bibinfo{person}{{COBHAM}}.} \bibinfo{year}{2018}\natexlab{}.
\newblock \bibinfo{title}{{GR712RC Dual-Core LEON3-FT SPARC V8 Processor Data
  Sheet}}.
\newblock
  \bibinfo{howpublished}{\url{https://www.gaisler.com/doc/gr712rc-datasheet.pdf}
  (Accessed Oct 2019)}.
\newblock


\bibitem[\protect\citeauthoryear{Dennehy}{Dennehy}{[n.d.]}]%
        {dennehy}
\bibfield{author}{\bibinfo{person}{Cornelius Dennehy}.}
  \bibinfo{year}{[n.d.]}\natexlab{}.
\newblock \bibinfo{title}{A NASA GN\&C Viewpoint on On-Board Processing
  Challenges to Support Optical Navigation and Other GN\&C Critical Functions}.
\newblock
\newblock
\urldef\tempurl%
\url{https://indico.esa.int/event/225/contributions/4249/}
\showURL{%
\tempurl}


\bibitem[\protect\citeauthoryear{Drabbe and Drabbe}{Drabbe and Drabbe}{2008}]%
        {drabbe_drabbe_2008}
\bibfield{author}{\bibinfo{person}{Jacco Drabbe} {and} \bibinfo{person}{Jacco
  Drabbe}.} \bibinfo{year}{2008}\natexlab{}.
\newblock \bibinfo{title}{ECSS-E-ST-70-11C – Space segment operability}.
\newblock
\newblock
\urldef\tempurl%
\url{https://ecss.nl/standard/ecss-e-st-70-11c-space-segment-operability/}
\showURL{%
\tempurl}


\bibitem[\protect\citeauthoryear{Dutta, Kushner, Jha, Sankaranarayanan,
  Shankar, and Tiwari}{Dutta et~al\mbox{.}}{[n.d.]}]%
        {duttasherlock}
\bibfield{author}{\bibinfo{person}{Souradeep Dutta}, \bibinfo{person}{Taisa
  Kushner}, \bibinfo{person}{Susmit Jha}, \bibinfo{person}{Sriram
  Sankaranarayanan}, \bibinfo{person}{Natarajan Shankar}, {and}
  \bibinfo{person}{Ashish Tiwari}.} \bibinfo{year}{[n.d.]}\natexlab{}.
\newblock \showarticletitle{Sherlock: A Tool for Verification of Deep Neural
  Networks}.
\newblock  (\bibinfo{year}{[n.\,d.]}).
\newblock


\bibitem[\protect\citeauthoryear{Esposito}{Esposito}{2019}]%
        {esposito_2019}
\bibfield{author}{\bibinfo{person}{Marco Esposito}.}
  \bibinfo{year}{2019}\natexlab{}.
\newblock \bibinfo{title}{CloudScout: In Orbit Demonstration of Machine
  Learning applied on hyperspectral and multispectral thermal imaging}.
\newblock
\newblock
\urldef\tempurl%
\url{https://indico.esa.int/event/225/timetable/#20190225.detailed}
\showURL{%
\tempurl}


\bibitem[\protect\citeauthoryear{Estlin, Bornstein, Gaines, Anderson, Thompson,
  Burl, Casta{\~n}o, and Judd}{Estlin et~al\mbox{.}}{2012}]%
        {estlin2012aegis}
\bibfield{author}{\bibinfo{person}{Tara~A Estlin}, \bibinfo{person}{Benjamin~J
  Bornstein}, \bibinfo{person}{Daniel~M Gaines}, \bibinfo{person}{Robert~C
  Anderson}, \bibinfo{person}{David~R Thompson}, \bibinfo{person}{Michael
  Burl}, \bibinfo{person}{Rebecca Casta{\~n}o}, {and} \bibinfo{person}{Michele
  Judd}.} \bibinfo{year}{2012}\natexlab{}.
\newblock \showarticletitle{Aegis automated science targeting for the mer
  opportunity rover}.
\newblock \bibinfo{journal}{\emph{ACM Transactions on Intelligent Systems and
  Technology (TIST)}} \bibinfo{volume}{3}, \bibinfo{number}{3}
  (\bibinfo{year}{2012}), \bibinfo{pages}{50}.
\newblock


\bibitem[\protect\citeauthoryear{Evers}{Evers}{2019}]%
        {evers2019deep}
\bibfield{author}{\bibinfo{person}{Nick Evers}.}
  \bibinfo{year}{2019}\natexlab{}.
\newblock \bibinfo{title}{{Deep learning in Space}}.
\newblock
  \bibinfo{howpublished}{\url{https://towardsdatascience.com/deep-learning-in-space-964566f09dcd}
  (Accessed Oct 2019)}.
\newblock


\bibitem[\protect\citeauthoryear{Ferraguto, Wittrock, Barrenscheen, Paakko, and
  Sipinen}{Ferraguto et~al\mbox{.}}{2008}]%
        {ferraguto2008board}
\bibfield{author}{\bibinfo{person}{Massimo Ferraguto}, \bibinfo{person}{Tim
  Wittrock}, \bibinfo{person}{Mark Barrenscheen}, \bibinfo{person}{Matti
  Paakko}, {and} \bibinfo{person}{Ville Sipinen}.}
  \bibinfo{year}{2008}\natexlab{}.
\newblock \showarticletitle{The on-board control procedures subsystem for the
  Herschel and Planck satellites}. In \bibinfo{booktitle}{\emph{2008 32nd
  Annual IEEE International Computer Software and Applications Conference}}.
  IEEE, \bibinfo{pages}{1366--1371}.
\newblock


\bibitem[\protect\citeauthoryear{Ferreira, Paffenroth, Wyglinski, Hackett,
  Bilen, Reinhart, and Mortensen}{Ferreira et~al\mbox{.}}{2019}]%
        {ferreira2019reinforcement}
\bibfield{author}{\bibinfo{person}{Paulo Victor~R Ferreira},
  \bibinfo{person}{Randy Paffenroth}, \bibinfo{person}{Alexander~M Wyglinski},
  \bibinfo{person}{Timothy~M Hackett}, \bibinfo{person}{Sven~G Bilen},
  \bibinfo{person}{Richard~C Reinhart}, {and} \bibinfo{person}{Dale~J
  Mortensen}.} \bibinfo{year}{2019}\natexlab{}.
\newblock \showarticletitle{Reinforcement Learning for Satellite
  Communications: From LEO to Deep Space Operations}.
\newblock \bibinfo{journal}{\emph{IEEE Communications Magazine}}
  \bibinfo{volume}{57}, \bibinfo{number}{5} (\bibinfo{year}{2019}),
  \bibinfo{pages}{70--75}.
\newblock


\bibitem[\protect\citeauthoryear{Gao, Samperio, Shala, and Cheng}{Gao
  et~al\mbox{.}}{2012}]%
        {gao2012modular}
\bibfield{author}{\bibinfo{person}{Yang Gao}, \bibinfo{person}{Renato
  Samperio}, \bibinfo{person}{Karin Shala}, {and} \bibinfo{person}{Y Cheng}.}
  \bibinfo{year}{2012}\natexlab{}.
\newblock \showarticletitle{Modular design for planetary rover autonomous
  navigation software using ROS}.
\newblock \bibinfo{journal}{\emph{Acta Futura}} \bibinfo{number}{5}
  (\bibinfo{year}{2012}), \bibinfo{pages}{9--16}.
\newblock


\bibitem[\protect\citeauthoryear{George and Wilson}{George and Wilson}{2018}]%
        {george2018onboard}
\bibfield{author}{\bibinfo{person}{Alan~D George} {and}
  \bibinfo{person}{Christopher~M Wilson}.} \bibinfo{year}{2018}\natexlab{}.
\newblock \showarticletitle{Onboard processing with hybrid and reconfigurable
  computing on small satellites}.
\newblock \bibinfo{journal}{\emph{Proc. IEEE}} \bibinfo{volume}{106},
  \bibinfo{number}{3} (\bibinfo{year}{2018}), \bibinfo{pages}{458--470}.
\newblock


\bibitem[\protect\citeauthoryear{Gretok, Kain, and George}{Gretok
  et~al\mbox{.}}{2019}]%
        {gretok2019comparative}
\bibfield{author}{\bibinfo{person}{Evan~W Gretok}, \bibinfo{person}{Evan~T
  Kain}, {and} \bibinfo{person}{Alan~D George}.}
  \bibinfo{year}{2019}\natexlab{}.
\newblock \showarticletitle{Comparative Benchmarking Analysis of
  Next-Generation Space Processors}. In \bibinfo{booktitle}{\emph{2019 IEEE
  Aerospace Conference}}. IEEE, \bibinfo{pages}{1--16}.
\newblock


\bibitem[\protect\citeauthoryear{Guirado, Tabik, Rivas, Alcaraz-Segura, and
  Herrera}{Guirado et~al\mbox{.}}{2019}]%
        {guirado2019whale}
\bibfield{author}{\bibinfo{person}{Emilio Guirado}, \bibinfo{person}{Siham
  Tabik}, \bibinfo{person}{Marga~L Rivas}, \bibinfo{person}{Domingo
  Alcaraz-Segura}, {and} \bibinfo{person}{Francisco Herrera}.}
  \bibinfo{year}{2019}\natexlab{}.
\newblock \showarticletitle{Whale counting in satellite and aerial images with
  deep learning}.
\newblock \bibinfo{journal}{\emph{Scientific reports}} \bibinfo{volume}{9},
  \bibinfo{number}{1} (\bibinfo{year}{2019}), \bibinfo{pages}{1--12}.
\newblock


\bibitem[\protect\citeauthoryear{{ISIS Space}}{{ISIS Space}}{2018}]%
        {sbandantenna}
\bibfield{author}{\bibinfo{person}{{ISIS Space}}.}
  \bibinfo{year}{2018}\natexlab{}.
\newblock \bibinfo{title}{{ISIS High Data Rate S-Band Transmitter}}.
\newblock
  \bibinfo{howpublished}{\url{https://www.isispace.nl/product/isis-txs-s-band-transmitter/}
  (Accessed Oct 2019)}.
\newblock


\bibitem[\protect\citeauthoryear{Jeppesen, Jacobsen, Inceoglu, and
  Toftegaard}{Jeppesen et~al\mbox{.}}{2019}]%
        {jeppesen2019cloud}
\bibfield{author}{\bibinfo{person}{Jacob~H{\o}xbroe Jeppesen},
  \bibinfo{person}{Rune~Hylsberg Jacobsen}, \bibinfo{person}{Fadil Inceoglu},
  {and} \bibinfo{person}{Thomas~Skj{\o}deberg Toftegaard}.}
  \bibinfo{year}{2019}\natexlab{}.
\newblock \showarticletitle{A cloud detection algorithm for satellite imagery
  based on deep learning}.
\newblock \bibinfo{journal}{\emph{Remote Sensing of Environment}}
  \bibinfo{volume}{229} (\bibinfo{year}{2019}), \bibinfo{pages}{247--259}.
\newblock


\bibitem[\protect\citeauthoryear{Lewis, Cote, and Tatnall}{Lewis
  et~al\mbox{.}}{1997}]%
        {lewis1997determination}
\bibfield{author}{\bibinfo{person}{HG Lewis}, \bibinfo{person}{S Cote}, {and}
  \bibinfo{person}{ARL Tatnall}.} \bibinfo{year}{1997}\natexlab{}.
\newblock \showarticletitle{Determination of spatial and temporal
  characteristics as an aid to neural network cloud classification}.
\newblock \bibinfo{journal}{\emph{International Journal of Remote Sensing}}
  \bibinfo{volume}{18}, \bibinfo{number}{4} (\bibinfo{year}{1997}),
  \bibinfo{pages}{899--915}.
\newblock


\bibitem[\protect\citeauthoryear{Liberis and Lane}{Liberis and Lane}{2019}]%
        {liberis2019neural}
\bibfield{author}{\bibinfo{person}{Edgar Liberis} {and}
  \bibinfo{person}{Nicholas~D Lane}.} \bibinfo{year}{2019}\natexlab{}.
\newblock \showarticletitle{Neural networks on microcontrollers: saving memory
  at inference via operator reordering}.
\newblock \bibinfo{journal}{\emph{arXiv preprint arXiv:1910.05110}}
  (\bibinfo{year}{2019}).
\newblock


\bibitem[\protect\citeauthoryear{Lorenz, Olds, May, Mario, Perry, Palmer, and
  Daly}{Lorenz et~al\mbox{.}}{2017}]%
        {lorenz2017lessons}
\bibfield{author}{\bibinfo{person}{David~A Lorenz}, \bibinfo{person}{Ryan
  Olds}, \bibinfo{person}{Alexander May}, \bibinfo{person}{Courtney Mario},
  \bibinfo{person}{Mark~E Perry}, \bibinfo{person}{Eric~E Palmer}, {and}
  \bibinfo{person}{Michael Daly}.} \bibinfo{year}{2017}\natexlab{}.
\newblock \showarticletitle{Lessons learned from OSIRIS-Rex autonomous
  navigation using natural feature tracking}. In \bibinfo{booktitle}{\emph{2017
  IEEE Aerospace Conference}}. IEEE, \bibinfo{pages}{1--12}.
\newblock


\bibitem[\protect\citeauthoryear{Manning, Langerman, Ramesh, Gretok, Wilson,
  George, MacKinnon, and Crum}{Manning et~al\mbox{.}}{2018}]%
        {manning2018machine}
\bibfield{author}{\bibinfo{person}{Jacob Manning}, \bibinfo{person}{David
  Langerman}, \bibinfo{person}{Barath Ramesh}, \bibinfo{person}{Evan Gretok},
  \bibinfo{person}{Christopher Wilson}, \bibinfo{person}{Alan George},
  \bibinfo{person}{James MacKinnon}, {and} \bibinfo{person}{Gary Crum}.}
  \bibinfo{year}{2018}\natexlab{}.
\newblock \showarticletitle{Machine-learning space applications on smallsat
  platforms with tensorflow}. In \bibinfo{booktitle}{\emph{Proceedings of the
  32nd Annual AIAA/USU Conference on Small Satellites, Logan, UT, USA}}.
  \bibinfo{pages}{4--9}.
\newblock


\bibitem[\protect\citeauthoryear{{Mohajerani} and {Saeedi}}{{Mohajerani} and
  {Saeedi}}{2019}]%
        {38-cloud-1}
\bibfield{author}{\bibinfo{person}{S. {Mohajerani}} {and} \bibinfo{person}{P.
  {Saeedi}}.} \bibinfo{year}{2019}\natexlab{}.
\newblock \showarticletitle{Cloud-Net: An End-To-End Cloud Detection Algorithm
  for Landsat 8 Imagery}. In \bibinfo{booktitle}{\emph{IGARSS 2019 - 2019 IEEE
  International Geoscience and Remote Sensing Symposium}}.
  \bibinfo{pages}{1029--1032}.
\newblock
\showISSN{2153-6996}
\urldef\tempurl%
\url{https://doi.org/10.1109/IGARSS.2019.8898776}
\showDOI{\tempurl}


\bibitem[\protect\citeauthoryear{Moreira}{Moreira}{2019}]%
        {sarlecture}
\bibfield{author}{\bibinfo{person}{Alberto Moreira}.}
  \bibinfo{year}{2019}\natexlab{}.
\newblock \bibinfo{title}{{Synthetic Aperture Radar (SAR): Principles and
  Applications}}.
\newblock
  \bibinfo{howpublished}{\url{https://earth.esa.int/documents/10174/642943/6-LTC2013-SAR-Moreira.pdf}
  (Accessed Oct 2019)}.
\newblock


\bibitem[\protect\citeauthoryear{Nanjangud, Blacker, Bandyopadhyay, and
  Gao}{Nanjangud et~al\mbox{.}}{2018}]%
        {nanjangud2018robotics}
\bibfield{author}{\bibinfo{person}{Angadh Nanjangud}, \bibinfo{person}{Peter
  Blacker}, \bibinfo{person}{Saptarshi Bandyopadhyay}, {and}
  \bibinfo{person}{Yang Gao}.} \bibinfo{year}{2018}\natexlab{}.
\newblock \showarticletitle{Robotics and AI-enabled on-orbit operations with
  future generation of small satellites}.
\newblock \bibinfo{journal}{\emph{Proc. IEEE}} \bibinfo{volume}{106},
  \bibinfo{number}{3} (\bibinfo{year}{2018}), \bibinfo{pages}{429--439}.
\newblock


\bibitem[\protect\citeauthoryear{{NASA}}{{NASA}}{2018}]%
        {nasasmallsat}
\bibfield{author}{\bibinfo{person}{{NASA}}.} \bibinfo{year}{2018}\natexlab{}.
\newblock \bibinfo{title}{State of the Art Small Spacecraft Technology}.
\newblock
  \bibinfo{howpublished}{\url{https://www.nasa.gov/sites/default/files/atoms/files/soa2018_final_doc.pdf}
  (Accessed Oct 2019)}.
\newblock


\bibitem[\protect\citeauthoryear{Notebaert}{Notebaert}{[n.d.]}]%
        {notebaert}
\bibfield{author}{\bibinfo{person}{Olivier Notebaert}.}
  \bibinfo{year}{[n.d.]}\natexlab{}.
\newblock \bibinfo{title}{On-Board Payload Data Processing requirements and
  ...}
\newblock
\newblock
\urldef\tempurl%
\url{https://indico.esa.int/event/225/contributions/4298/attachments/3359/4397/OBDP2019-S01-05-Airbus_Notebaert_On-Board_Payload_Data_Processing_Requirements_and_Technology_Trends.pdf}
\showURL{%
\tempurl}


\bibitem[\protect\citeauthoryear{Ort{\'\i}z-G{\'o}mez, Rodr{\'\i}guez-Osorio,
  Salas-Natera, Landeros-Ayala, Tarchi, and
  Vanelli-Coralli}{Ort{\'\i}z-G{\'o}mez et~al\mbox{.}}{[n.d.]}]%
        {ortizuse}
\bibfield{author}{\bibinfo{person}{Flor~G Ort{\'\i}z-G{\'o}mez},
  \bibinfo{person}{Ram{\'o}n~Mart{\'\i}nez Rodr{\'\i}guez-Osorio},
  \bibinfo{person}{Miguel~A Salas-Natera}, \bibinfo{person}{Salvador
  Landeros-Ayala}, \bibinfo{person}{Daniele Tarchi}, {and}
  \bibinfo{person}{Alessandro Vanelli-Coralli}.}
  \bibinfo{year}{[n.d.]}\natexlab{}.
\newblock \showarticletitle{ON THE USE OF NEURAL NETWORKS FOR FLEXIBLE PAYLOAD
  MANAGEMENT IN VHTS SYSTEMS}.
\newblock  (\bibinfo{year}{[n.\,d.]}).
\newblock


\bibitem[\protect\citeauthoryear{Paoletti, Haut, Plaza, and Plaza}{Paoletti
  et~al\mbox{.}}{2019}]%
        {paoletti2019deep}
\bibfield{author}{\bibinfo{person}{ME Paoletti}, \bibinfo{person}{JM Haut},
  \bibinfo{person}{J Plaza}, {and} \bibinfo{person}{A Plaza}.}
  \bibinfo{year}{2019}\natexlab{}.
\newblock \showarticletitle{Deep learning classifiers for hyperspectral
  imaging: A review}.
\newblock \bibinfo{journal}{\emph{ISPRS Journal of Photogrammetry and Remote
  Sensing}}  \bibinfo{volume}{158} (\bibinfo{year}{2019}),
  \bibinfo{pages}{279--317}.
\newblock


\bibitem[\protect\citeauthoryear{Qian, Bergeron, Cunningham, Gagnon, and
  Hollinger}{Qian et~al\mbox{.}}{2006}]%
        {qian2006near}
\bibfield{author}{\bibinfo{person}{Shen-En Qian}, \bibinfo{person}{Martin
  Bergeron}, \bibinfo{person}{Ian Cunningham}, \bibinfo{person}{Luc Gagnon},
  {and} \bibinfo{person}{Allan Hollinger}.} \bibinfo{year}{2006}\natexlab{}.
\newblock \showarticletitle{Near lossless data compression onboard a
  hyperspectral satellite}.
\newblock \bibinfo{journal}{\emph{IEEE Trans. Aerospace Electron. Systems}}
  \bibinfo{volume}{42}, \bibinfo{number}{3} (\bibinfo{year}{2006}),
  \bibinfo{pages}{851--866}.
\newblock


\bibitem[\protect\citeauthoryear{Salem, Kafatos, El-Ghazawi, Gomez, and
  Yang}{Salem et~al\mbox{.}}{2001}]%
        {salem2001hyperspectral}
\bibfield{author}{\bibinfo{person}{Foudan Salem}, \bibinfo{person}{Menas
  Kafatos}, \bibinfo{person}{Tarek El-Ghazawi}, \bibinfo{person}{Richard
  Gomez}, {and} \bibinfo{person}{Ruixin Yang}.}
  \bibinfo{year}{2001}\natexlab{}.
\newblock \showarticletitle{Hyperspectral image analysis for oil spill
  detection}. In \bibinfo{booktitle}{\emph{Summaries of NASA/JPL Airborne Earth
  Science Workshop, Pasadena, CA}}. \bibinfo{pages}{5--9}.
\newblock


\bibitem[\protect\citeauthoryear{{Satellite Imaging Corporation}}{{Satellite
  Imaging Corporation}}{2019}]%
        {wv4sat}
\bibfield{author}{\bibinfo{person}{{Satellite Imaging Corporation}}.}
  \bibinfo{year}{2019}\natexlab{}.
\newblock \bibinfo{title}{{WorldView-4 Satellite Sensor}}.
\newblock
  \bibinfo{howpublished}{\url{https://www.satimagingcorp.com/satellite-sensors/geoeye-2/}
  (Accessed Oct 2019)}.
\newblock


\bibitem[\protect\citeauthoryear{Schartel}{Schartel}{2017}]%
        {schartel2017increasing}
\bibfield{author}{\bibinfo{person}{Andreas Schartel}.}
  \bibinfo{year}{2017}\natexlab{}.
\newblock \bibinfo{title}{Increasing Spacecraft Autonomy through Embedded
  Neural Networks for Semantic Image Analysis}.
\newblock
\newblock


\bibitem[\protect\citeauthoryear{Sze, Chen, Yang, and Emer}{Sze
  et~al\mbox{.}}{2017}]%
        {sze2017efficient}
\bibfield{author}{\bibinfo{person}{Vivienne Sze}, \bibinfo{person}{Yu-Hsin
  Chen}, \bibinfo{person}{Tien-Ju Yang}, {and} \bibinfo{person}{Joel~S Emer}.}
  \bibinfo{year}{2017}\natexlab{}.
\newblock \showarticletitle{Efficient processing of deep neural networks: A
  tutorial and survey}.
\newblock \bibinfo{journal}{\emph{Proc. IEEE}} \bibinfo{volume}{105},
  \bibinfo{number}{12} (\bibinfo{year}{2017}), \bibinfo{pages}{2295--2329}.
\newblock


\bibitem[\protect\citeauthoryear{Torelli}{Torelli}{2019}]%
        {torelli_2019}
\bibfield{author}{\bibinfo{person}{Felice Torelli}.}
  \bibinfo{year}{2019}\natexlab{}.
\newblock \bibinfo{title}{Common DPU and Basic SW for JUICE instruments}.
\newblock
\newblock
\urldef\tempurl%
\url{https://indico.esa.int/event/225/contributions/3688/}
\showURL{%
\tempurl}


\bibitem[\protect\citeauthoryear{Trivedi and Mehta}{Trivedi and Mehta}{2016}]%
        {trivedi2016survey}
\bibfield{author}{\bibinfo{person}{Rakesh Trivedi} {and}
  \bibinfo{person}{Usha~S Mehta}.} \bibinfo{year}{2016}\natexlab{}.
\newblock \showarticletitle{A survey of radiation hardening by design (rhbd)
  techniques for electronic systems for space application}.
\newblock \bibinfo{journal}{\emph{International Journal of Electronics and
  Communication Engineering Technology (IJECET)}} \bibinfo{volume}{7},
  \bibinfo{number}{1} (\bibinfo{year}{2016}), \bibinfo{pages}{75}.
\newblock


\bibitem[\protect\citeauthoryear{Usgs}{Usgs}{[n.d.]}]%
        {usgs}
\bibfield{author}{\bibinfo{person}{Usgs}.} \bibinfo{year}{[n.d.]}\natexlab{}.
\newblock \bibinfo{title}{Home}.
\newblock
\newblock
\urldef\tempurl%
\url{https://earthexplorer.usgs.gov/}
\showURL{%
\tempurl}


\bibitem[\protect\citeauthoryear{Valsesia and Magli}{Valsesia and
  Magli}{2019}]%
        {diego2019image}
\bibfield{author}{\bibinfo{person}{Diego Valsesia} {and}
  \bibinfo{person}{Enrico Magli}.} \bibinfo{year}{2019}\natexlab{}.
\newblock \bibinfo{title}{{Image dequantization for hyperspectral lossy
  compression with convolutional neural networks}}.
\newblock
  \bibinfo{howpublished}{\url{https://indico.esa.int/event/225/contributions/245487/}
  (Accessed Oct 2019)}.
\newblock


\bibitem[\protect\citeauthoryear{Wang and Patel}{Wang and Patel}{2018}]%
        {wang2018generating}
\bibfield{author}{\bibinfo{person}{Puyang Wang} {and} \bibinfo{person}{Vishal~M
  Patel}.} \bibinfo{year}{2018}\natexlab{}.
\newblock \showarticletitle{Generating high quality visible images from SAR
  images using CNNs}. In \bibinfo{booktitle}{\emph{2018 IEEE Radar Conference
  (RadarConf18)}}. IEEE, \bibinfo{pages}{0570--0575}.
\newblock


\bibitem[\protect\citeauthoryear{Wang, Ding, Li, Yuan, Liao, Ma, Yuan, Qian,
  Tang, Qiu, et~al\mbox{.}}{Wang et~al\mbox{.}}{2018}]%
        {wang2018towards}
\bibfield{author}{\bibinfo{person}{Yanzhi Wang}, \bibinfo{person}{Caiwen Ding},
  \bibinfo{person}{Zhe Li}, \bibinfo{person}{Geng Yuan}, \bibinfo{person}{Siyu
  Liao}, \bibinfo{person}{Xiaolong Ma}, \bibinfo{person}{Bo Yuan},
  \bibinfo{person}{Xuehai Qian}, \bibinfo{person}{Jian Tang},
  \bibinfo{person}{Qinru Qiu}, {et~al\mbox{.}}}
  \bibinfo{year}{2018}\natexlab{}.
\newblock \showarticletitle{Towards ultra-high performance and energy
  efficiency of deep learning systems: an algorithm-hardware co-optimization
  framework}. In \bibinfo{booktitle}{\emph{Thirty-Second AAAI Conference on
  Artificial Intelligence}}.
\newblock


\bibitem[\protect\citeauthoryear{Wimmers, Velden, and Cossuth}{Wimmers
  et~al\mbox{.}}{2019}]%
        {wimmers2019using}
\bibfield{author}{\bibinfo{person}{Anthony Wimmers},
  \bibinfo{person}{Christopher Velden}, {and} \bibinfo{person}{Joshua~H
  Cossuth}.} \bibinfo{year}{2019}\natexlab{}.
\newblock \showarticletitle{Using deep learning to estimate tropical cyclone
  intensity from satellite passive microwave imagery}.
\newblock \bibinfo{journal}{\emph{Monthly Weather Review}}
  \bibinfo{volume}{147}, \bibinfo{number}{6} (\bibinfo{year}{2019}).
\newblock


\bibitem[\protect\citeauthoryear{Xiang, Musau, Wild, Lopez, Hamilton, Yang,
  Rosenfeld, and Johnson}{Xiang et~al\mbox{.}}{2018}]%
        {xiang2018verification}
\bibfield{author}{\bibinfo{person}{Weiming Xiang}, \bibinfo{person}{Patrick
  Musau}, \bibinfo{person}{Ayana~A Wild}, \bibinfo{person}{Diego~Manzanas
  Lopez}, \bibinfo{person}{Nathaniel Hamilton}, \bibinfo{person}{Xiaodong
  Yang}, \bibinfo{person}{Joel Rosenfeld}, {and} \bibinfo{person}{Taylor~T
  Johnson}.} \bibinfo{year}{2018}\natexlab{}.
\newblock \showarticletitle{Verification for machine learning, autonomy, and
  neural networks survey}.
\newblock \bibinfo{journal}{\emph{arXiv preprint arXiv:1810.01989}}
  (\bibinfo{year}{2018}).
\newblock


\end{thebibliography}

	}
	
\end{document}